\documentclass{iaus}
\usepackage{graphicx}

\title[Search for star-planet interaction] 
{Search for star-planet interaction}

\author[Krej\v{c}ov\'a, T., Budaj, J., Koza, J.]   
{Tereza Krej\v{c}ov\'a$^1$,
 J\'an Budaj$^2$ \and J\'ulius Koza$^2$}

\affiliation{$^1$Dept. of Theoretical Physics and Astrophysics,
Masaryk University,
\\Brno, Czech Republic, email: {\tt terak@physics.muni.cz} \\[\affilskip]
$^2$Astronomical Institute, Tatransk\'a Lomnica, Slovak Republic, email: {\tt budaj@ta3.sk}}

\pubyear{2008}
\volume{xxx}  
\pagerange{119--126}
\jname{Title of your IAU Symposium}
\editors{A.C. Editor, B.D. Editor \& C.E. Editor, eds.}
\begin{document}

\maketitle

\begin{abstract}
We analyse the chromospherical activity of stars with extrasolar planets
and search for a possible correlation between the equivalent width of the
core of Ca\,II\,K line and orbital parameters of the planet.
We found a statistically significant evidence that the equivalent
width of the Ca\,II\,K line reversal, which originates in the stellar 
chromosphere depends on the orbital period $P_{\mathrm{orb}}$ of 
the exoplanet. Planets orbiting stars with $T_{\mathrm{eff}}<5\,500$\,K 
and with $P_{\mathrm{orb}}<20$\,days generally have much stronger emission 
than planets at similar temperatures but at longer orbital periods.
$P_{\mathrm{orb}}=20$\,days marks a sudden change in behaviour, which
might be associated with a qualitative change in the star-planet
interaction.
\keywords{Ca\,II\,K line, exoplanet, star-planet interaction.}
\end{abstract}

\firstsection 

\section{Introduction}
The question of possible existence of star-planet interaction is currently studied in many ways. 
Based on the observations in the optical region \cite[Shkolnik et al. (2005, 2008)]{Shkolnik08}
discovered the
planetary induced variability in the cores of
Ca II
H \& K, H$\alpha$ and Ca\,II IR triplet in a few planet hosting stars. 
\cite[Knutson et al. (2010)]{Knutson10} found a correlation between 
the chromospheric activity of the star and presence of the
stratosphere on the planet. Consequently, \cite[Hartman (2010)]{Hartman10} 
found a correlation between the surface gravity of Hot Jupiters and 
the stellar activity. Recently \cite[Canto Martins et al. (2011)]{Martins11} 
searched for correlation between planetary parameters and 
the $\log R'_{\mathrm{HK}}$ parameter but didn't reveal
any convincing proof for such a phenomenon.

\section{Observation \& Statistical Analysis}
We used FEROS instrument on 2.2 ESO/MPG telescope to obtain spectra
of several stars (HD\,179949, HD\,212301, HD\,149143 and Wasp-18) 
with close-in exoplanet.
We also used the publicly available spectra from HIRES spectrograph archive.
Subsequently we measured the equivalent width of the central reversal 
in the core of Ca\,II\,K. 

In the first case we divided our data sample into two groups according 
to the semi-major axis ($a\leq0.15$ and $a>0.15$\,AU). Figure~1 (left-top) 
shows the dependence of equivalent width on the effective temperature of 
the star. Subsequently, we performed two statistical tests -- Student's-t 
test and Kolmogorov-Smirnov test 
to determine whether the two groups originate from the same population. 
The resulting probability is a function of temperature and is plotted in 
the lower part of Figure~1. The tests show that the difference between 
the two samples is significant for $T_{\mathrm{eff}}\leq 5\,500$\,K. 
It means that stars with lower temperature and with planets on closer orbits 
show more activity as measured in the core of Ca\,II\,K line.

In the second case we group the data according to the effective 
temperature of the parent star ($T_{\mathrm{eff}}\leq 5\,500$\,K 
and $T_{\mathrm{eff}}>5\,500$\,K) and plot the equivalent width of
the Ca\,II\,K line reversal as a function of the orbital period
(Figure~1--right).

\begin{figure}[!ht]
\begin{minipage}[htb]{0.47\linewidth}
\centering
\includegraphics[width=6.5cm]{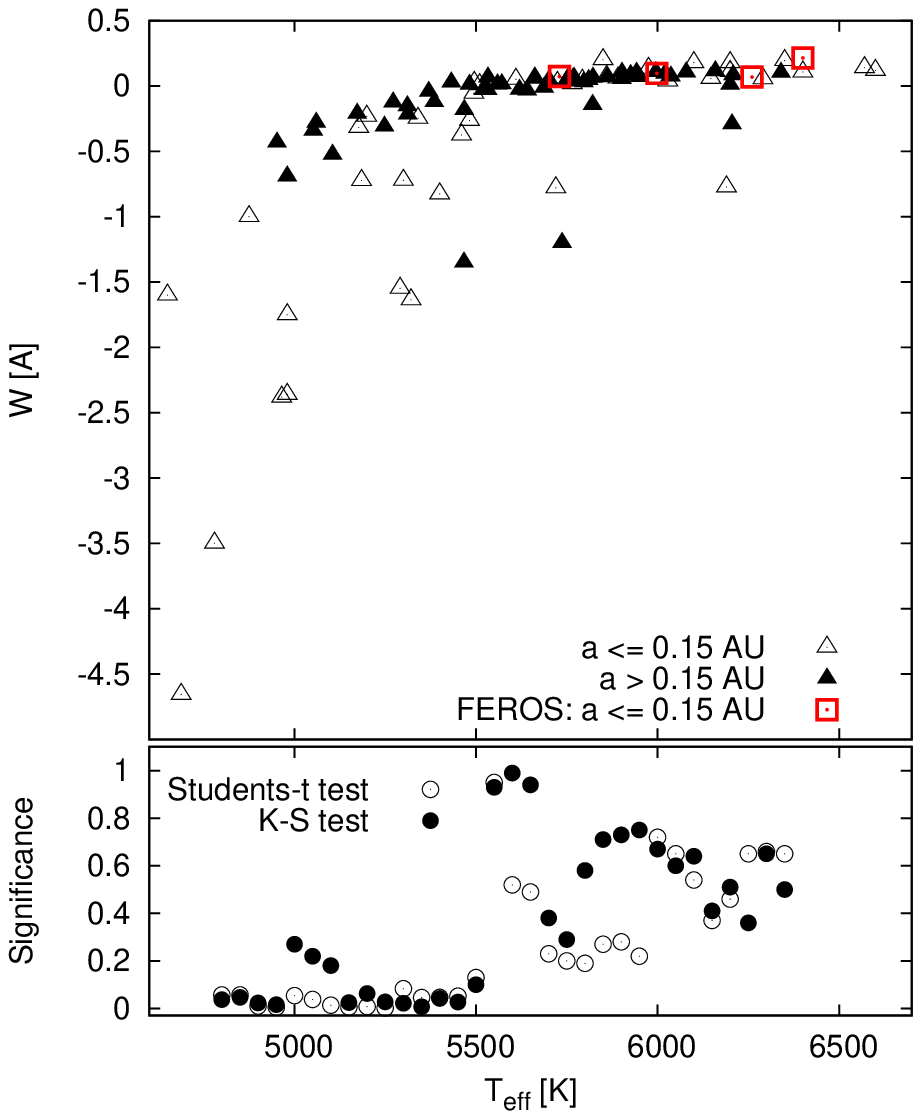}
\end{minipage}
\begin{minipage}[htb]{0.47\linewidth}
\centering
\includegraphics[width=7cm]{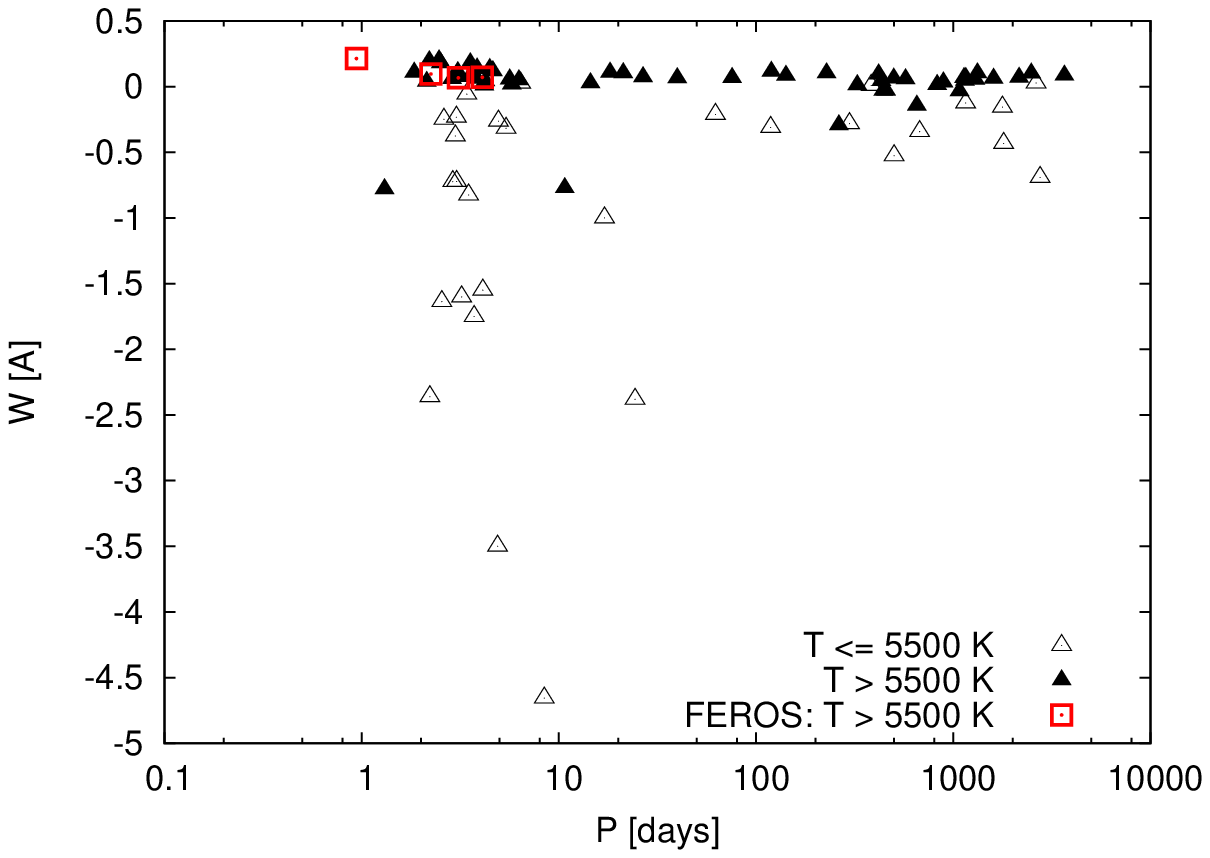}
\end{minipage}
\caption{
{\bf Left} {\it Top}: Dependence of the equivalent 
width of Ca\,II\,K reversal on the temperature of the parent star. 
Empty triangles 
are exoplanetary systems with $a\leq 0.15$\,AU, full triangles are 
systems with $a>0.15$\,AU. {\it Bottom}: Statistical Student's-t test (empty circles) and 
Kolmogorov-Smirnov test (full circles). Red squares are data from FEROS. {\bf Right} Dependence of the equivalent width of Ca\,II\,K on the orbital
period. Empty triangles are exoplanetary systems with $T\leq 5\,500$\,K, full triangles are systems with $T>5\,500$\,K and red squares are data from
FEROS. 
}
\label{f1}
\end{figure}

\noindent \textbf{Acknowledgements } This work has been supported by grant GA \v{C}R GD205/08/H005, VEGA 2/0078/10, VEGA 2/0074/09, VEGA
2/0094/11
and the National scholarship programme of Slovak Republic. This research has
made use of the Keck Observatory Archive (KOA), which is operated by the
W.\,M.\,Keck Observatory and the NASA Exoplanet Science
Institute (NExScI),
under contract with the National Aeronautics and Space Administration. We want to thank Tom\'a\v{s} Henych for fruitful discussion.


\begin{thebibliography}{}

\bibitem[Canto Martins et al. (2011)]{Martins11}
{Canto Martins, B.L., Das Chagas, M.L., Alves, S., et al.} 2011,
\textit{A\&A}, 530, A73 

\bibitem[Hartman (2010)]{Hartman10}
{Hartman, J.D.} 2010, 
\textit{ApJ}, 717, L138

\bibitem[Knutson et al. (2010)]{Knutson10}
{Knutson, H.A., Howard, A.W., \& Isaacson, H.} 2010, 
\textit{ApJ}, 720, 1569


\bibitem[Shkolnik et al. (2008)]{Shkolnik08}
{Shkolnik, E., Bohlender, D.A., Walker, G.A.H., \& Collier Cameron, A.} 2008, 
\textit{ApJ}, 676, 628

\bibitem[Shkolnik et al. (2005)]{Shkolnik05}
{Shkolnik, E., Walker, G.A.H., Bohlender, D.A., Gu, P., \& Kurster, M.} 2005,
\textit{ApJ}, 622, 1075



\end{thebibliography}
\end{document}